\documentclass[prl,aps,twocolumn,showpacs,superscriptaddress,amsmath,amssymb,floatfix,10pt]{revtex4-1}
\usepackage{graphicx}

\newcommand{\be}{\begin{equation}}
\newcommand{\ee}{\end{equation}}

\newcommand{\mume}{\mu \rm m}
\newcommand{\mev}{\rm meV}

\newcommand{\re}{\textrm{Re}}

\newcommand{\eq}[1]{(\ref{eq:#1})}

\begin{document}
\title{Energy relaxation in the Gross-Pitaevskii equation }
\author{Michiel Wouters}
\affiliation{Institute of Theoretical Physics, Ecole Polytechnique F\'ed\'erale de Lausanne (EPFL), CH-1015 Lausanne, Switzerland}
\author{Vincenzo Savona}
\affiliation{Institute of Theoretical Physics, Ecole Polytechnique F\'ed\'erale de Lausanne (EPFL), CH-1015 Lausanne, Switzerland}
\begin{abstract}
We introduce a dissipation term in the Gross-Pitaevskii equation that describes the stimulated relaxation of condensed bosons due to scattering with a different type of particles. This situation applies to Bose-Einstein condensates of quasi-particles in the solid state, such as magnons and excitons. Our model is compatible with the phenomenology of superfluidity: supercurrents are stable up to a critical speed and decay when they are faster. 
\end{abstract}
\maketitle

The Gross-Pitaevskii equation (GPE) provides an excellent description of an isolated dilute Bose gas at low temperature. For example, the GPE accurately describes the density profiles and frequencies of elementary excitations of trapped ultracold atomic bose gases\cite{lpss}.

Systems featuring Bose-Einstein Condensation (BEC) of quasi-particles in the solid state, such as (exciton-) polaritons~\cite{polcond} and magnons~\cite{magnon} are however not so well isolated from their environment as the atomic gases. In semiconductor microcavities for example, the polaritons that form a Bose Einstein condensate undergo scattering with lattice phonons and with high energy excitons. These interactions allow the polaritons to relax toward lower energy.

The Boltzmann equation is the easiest way to model these scattering dynamics theoretically ~\cite{boltzpol}. The semiclassical approximation involved in this approach becomes however problematic when the polariton gas enters the condensed phase. Most importantly, supercurrents, that are a gradient of the phase of the polariton field, are missed. This precludes e.g. the description of vortices in polariton condensates~\cite{konstantinos}, that appear naturally in the GPE.

\begin{figure}[htbp]
\begin{center}
\includegraphics[width=0.9\columnwidth,angle=0,clip]{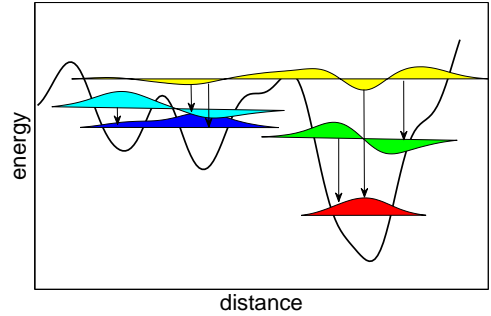}  
\end{center}
\caption{Schematic illustration of the relaxation of polaritons due to scattering with phonons. Several modes in a random potential are macroscopically occupied. Where they spatially overlap, relaxation takes locally place from the high to the low frequency modes, as indicated by the arrows.}
\label{fig:sketch}
\end{figure}

In order to allow for a simultaneous description supercurrents and of relaxation due to interactions with the environment, we will derive in this Letter a dissipation term that can be added to the GPE. This term models transitions between components of the GPE field at different frequencies (see Fig.~\ref{fig:sketch}). For specificity, we will use in the following the terminology of `polaritons' relaxing through `phonon scattering', but our formalism is applicable to any dilute Bose gas that dissipates energy into the environment.

An important requirement of the theory is compatibility with superfluidity. According to the Landau criterion~\cite{lpss}, the superflow cannot relax due to phonon scattering as long as the superfluid velocity is below the critical speed. A Boltzmann description that includes phonon scattering violates this requirement: a Bose gas with initially all the particles at finite momentum $k=k_c$ will relax to a state with the majority of particles in the ground state $k=0$. We will show below that our GPE based model for phonon relaxation is instead fully in agreement with the Landau criterion for stability of superflow slower than the critical speed. In addition, it provides a dynamical model for the dissipation of superflow when the speed exceeds the critical one.

A kinetic model formulated in terms of transitions between states characterized by their frequencies involves a coarse graining of the time, analogous to the coarse graining of space used in the semiclassical Boltzmann equation~\cite{boltzmann}. We will use the following  definition of the time dependent spectrum of the bose field $\psi(x,t)$:
\begin{equation}
\psi_\omega(x,t)=\frac{1}{T}\int_{t-T}^t e^{i\omega t'}\;\psi(x,t') dt',
\label{eq:ft}
\end{equation}
where $T$ is the coarse graining time step. The inverse transform reads
\begin{equation}
\psi(x,t)=\sum_\omega e^{-i\omega t} \psi_\omega(x,t).
\label{eq:ift}
\end{equation}
Spontaneous scattering is a quantum feature that is difficult to describe with a classical field model such as the GPE. We therefore restrict to the stimulated relaxation processes:
\begin{equation}
dn_\omega(x,t) = \, n_\omega(x,t)\, \sum_{\omega'} r(\omega,\omega') n_{\omega'}(x,t)\;dt,
\label{eq:dnomega}
\end{equation}
where $r(\omega,\omega')$ is the net scattering rate from the mode at frequency $\omega'$ to the mode at $\omega$. In order to conserve particle number, the relaxation rate should change sign under exchange of $\omega$ and $\omega'$: $r(\omega,\omega')=-r(\omega',\omega)$. 
The first term in the expansion of the rate as a function of the frequency difference is therefore
\begin{equation}
r(\omega,\omega')=\kappa \; (\omega'-\omega),
\label{eq:rw}
\end{equation}
where the relaxation constant $\kappa$ has the dimension of an inverse density.
Eq.~\eq{rw} is for example the form of the scattering rate obtained with the golden rule for the relaxation of quantum well excitons due to scattering with acoustic phonons~\cite{haug}. 

Under the change of the density~\eq{dnomega}, the wave function varies as
\begin{equation}
\psi_\omega(x,t)+d\psi_\omega(x,t)= \sqrt{\frac{n_\omega(x,t)+dn_\omega(x,t)}{n_\omega(x,t)}} \psi_\omega(x,t),
\label{eq:dpsiomega}
\end{equation}
where we used the fact that a stimulated relaxation process does not change the phase of $\psi$. Expanding Eq.~\eq{dpsiomega} to first order in $dn_\omega$, substituting Eqns.~\eq{dnomega} and~\eq{rw} and using the inverse transform~\eq{ift}, 
one obtains for the relaxation dynamics of the wave function
\begin{equation}
\frac{d\psi(x,t)}{dt} = \frac{\kappa \bar{n}(x,t)}{2}\left[
\bar{\mu}(x,t)-\frac{i \partial}{\partial t}
\right] \psi(x,t).
\label{eq:dpsi1}
\end{equation}
In the derivation, boundary terms in the partial integration that scale as $1/T$ were neglected, because we consider the limit of a large coarse graining time where $1/T$ is smaller than any other relevant frequency scale. In Eq.~\eq{dpsi1}, $\bar{n}$ and $\bar{\mu}$ are the time averaged density and chemical potential respectively:
\begin{eqnarray}
\bar{n}(x,t)&=&\frac{1}{T}\int_{t-T}^t |\psi(x,t')|^2 dt',\label{eq:Nb} \\
\bar{\mu}(x,t)&=&\frac{1}{\bar{n}(x,t)}\re\left[\frac{1}{T}\int_{t-T}^t \psi^*(x,t') 
 \frac{i\partial}{\partial t} \psi(x,t') dt'\right].
\label{eq:Eb}
\end{eqnarray}
The real part is taken in Eq.~\eq{Eb}, because it is readily shown with the Madelung transformation $\psi=\sqrt{n}\,e^{i\theta}$, that the imaginary part of the integral in Eq.~\eq{Eb} scales as $1/T$. Eq.~\eq{dpsi1} gives the term that was sought to describe the stimulated relaxation of a classical Bose field due to local scattering with phonon like particles and is the central result of this Letter.

The right hand side in Eq.~\eq{dpsi1} resembles the frequency dependent amplification term that was introduced in Ref.~\cite{nonres_superfl} to describe an energy dependent gain mechanism: $\partial \psi/\partial t=(P/\Omega_K)[\Omega_K-i\partial/\partial t] \psi$. Two main differences should be emphasized. First, here the relaxation term is proportional to the polariton density $\bar n$, whereas in Ref.~\cite{nonres_superfl}, the relaxation is proportional to the gain from the reservoir $P$. The second important difference is that the gain cutoff frequency $\Omega_K$ is replaced by the average frequency $\bar \mu$. As a consequence, particle loss and gain balance each other and there is no net gain in Eq.~\eq{dpsi1}. The modes with frequency $\omega>\bar \mu$ experience loss, where the ones with $\omega<\bar \mu$ are amplified. Phonon absorption is thus not included in our model. The relaxation~\eq{dpsi1} is due to the interaction with an environment that is effectively at zero temperature.

\begin{figure}[htbp]
\begin{center}
\includegraphics[width=\columnwidth,angle=0,clip]{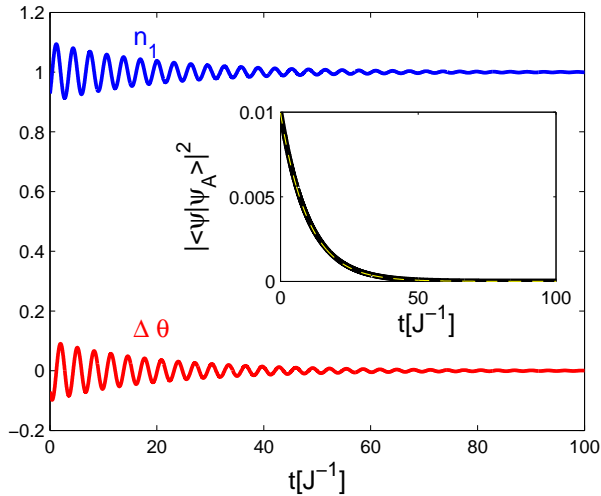}  
\end{center}
\caption{Dynamics of the relative phase ($\Delta \theta$) and density on the first well $(n_1)$ for two wells coupled by tunneling and dissipating through the emission of phonons. The initial condition was taken to be $\psi_{1,2}=e^{\pm i 0.05}$ Inset: The decay of the projection of the wave function on the excited state is excellently reproduced by the exponential decay $\exp(-2J \kappa n_1 t)$ (dashed line), where the phonon relaxation constant was taken $\kappa=0.05$ and the coarse graining time scale $T=20J^{-1}$.}
\label{fig:rabi}
\end{figure}

As a first application, we add the dissipation term~\eq{dpsi1} to the GPE for noninteracting particles in two coupled levels. In the context of polariton condensation, the coupled two-level system describes for example the situation of spatially distinct potential wells~\cite{josephson} or the polarization degree~\cite{shelykyreview}. The full dynamics is
\begin{equation}
i\frac{\partial}{\partial t}\psi_j(t)= -J \psi_{3-j}
+i\frac{\kappa \bar{n}_j(t)}{2} \left[
\bar{\mu}_j(t)- \frac{i \partial}{\partial t}
\right] \psi_j(t),
\label{eq:GPET}
\end{equation}
for $j=1,2$, where $J$ is the coupling energy. Fig.~\ref{fig:rabi} shows the time evolution of the densities in the two states, obtained from the numerical integration of Eq.~\eq{GPET}. The initial wave function is taken to have equal amplitudes on the two states and a small phase difference between them. Due to the dissipation, the Rabi oscillations are damped. Note that the time evolution~\eq{GPET} drives the system automatically into the ground state and that the eigenstates of the Hamiltonian were not explicitly needed to compute the phonon scattering rates.  The contribution of the excited state to the wave function is shown in the inset of Fig.~\ref{fig:rabi}. Its exponential decay is excellently reproduced with the rate obtained from Eq.~\eq{dnomega}. 

We have numerically checked that the dissipative dynamics conserves particle number in the limit of a small time step in the numerical integration. We correct the residual deviation by restoring the norm of the wave function after each application of the dissipation operator.

We now show that in the case of the interacting homogeneous Bose gas, our description of the dissipation is fully compatible with the phenomenology of persistent supercurrents. The Gross-Pitaevskii equation for the homogeneous interacting Bose gas supplemented with the phonon scattering term reads
\begin{equation}
i \frac{\partial}{\partial t} \psi= \left[-\frac{\nabla^2}{2m}+g|\psi|^2+i\frac{\kappa \bar{n}}{2} 
\left(
\bar{\mu}- \frac{i \partial}{\partial t}
\right)
\right] \psi.
\label{eq:GPEP}
\end{equation}
The usual steady state solution $\psi_0(x,t)=\sqrt{n_c}e^{-i\mu t + ik_cx}$ is still a solution of Eq.~\eq{GPEP} with $\mu=k_c^2/2m+gn_c$, where $k_c$ is the condensate momentum, $g$ the interaction constant and $n_c$ the condensate density.

The excitations on top of the condensate can be described by the wave function $\psi(x,t)=\psi_0(x,t)[1+ue^{i k x -i \omega(k) t}+v^*e^{-i k x +i \omega^*(k) t}]$. Linearizing the equations of motion in $u$ and $v$ is not more difficult than in the standard case, because to first order in $u$ and $v$, the average density $\bar{n}$ and frequency $\bar{\mu}$ do not change. In the small dissipation limit $\kappa n\ll 1$, the frequencies of the elementary excitations read : 
\begin{multline}
\omega(k)=\pm\sqrt{\omega^2_B(k)-(\kappa n/2)^2} + k k_c/m \\
-i\frac{\kappa n}{2}
\left[
\frac{k^2/2m+gn}{\omega_B(k)}
\right]
\left[\omega_B(k)\pm k k_c  \right].
\label{eq:omega}
\end{multline}
Note that the correction quadratic in $\kappa n$ to the usual Bogoliubov dispersion $\omega_B(k)=\sqrt{gnk^2/m+k^4/4m}$ under the square root on the first line in Eq.~\eq{omega} is necessary to satisfy the Goldstone theorem that requires at least one branch of elementary excitations to have exactly zero real and imaginary part of the frequency for $k\rightarrow 0$.

Importantly, for a condensate wave vector below the critical one $k_{\rm crit}={\rm min}[\omega_B(k)/k]$, the elementary excitations have a negative imaginary part and the condensate is dynamically stable against decay into a lower energy state. 
As soon as $k_c>k_{\rm crit}$, the imaginary part of the excitations becomes positive and the condensate becomes unstable with respect to decay toward a lower momentum. In agreement with the thermodynamical considerations made by Landau~\cite{lpss}, the phonon scattering is able to dissipate the supercurrent only when the condensate velocity is above the critical speed. In the present dynamical model~\eq{GPEP}, the energetic instability gives rise to a dynamical one, because the dissipative environment is explicitly included and describes the kinetics of the supercurrent's decay.

\begin{figure}[htbp]
\begin{center}
\includegraphics[width=\columnwidth,angle=0,clip]{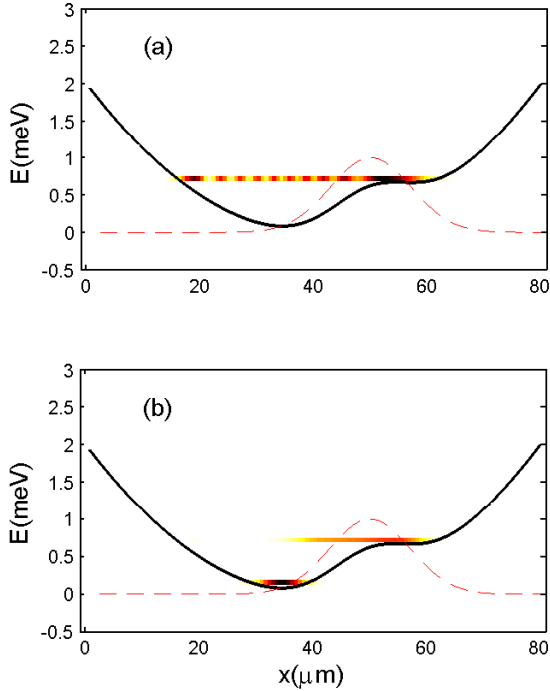}  
\end{center}
\caption{Relaxation of a polariton condensate that is nonresonantly created by a pump profile (dashed line) that is not centered at the minimum of the effective trapping potential (full line). \textit{Panel (a)}: In the absence of a relaxation mechanism ($\kappa=0$), the polaritons accelerate ballistically when moving toward the bottom of the potential: The energy resolved real space distribution shows a single frequency for the condensate. \textit{Panel (b)}: When the dissipation mechanism is present ($\kappa=0.05 \mume^2$), the polaritons relax to the bottom of the trap. The condensate frequency in the center of the trap is different from the condensate frequency in the pumping area. The red dashed line shows the excitation intensity $P$ and the full black line represents the effective potential $V_{\rm eff}$. Other parameters: $\hbar \gamma=0.05 \mev$, $m_{LP}/\hbar^2=\mev^{-1}\mume^{-2}$, $\hbar P=1 \mev \mume^{-2}$, $\hbar \gamma_R=0.5 \mev$, $\hbar R(n_R)=(1 \mev \mume^2)n_R $. }
\label{fig:snoke}
\end{figure}

As a last illustration, we describe the relaxation of polaritons in a harmonic trap as observed in Ref.~\cite{snoke}. When polaritons were created off center, relaxation toward the lower energy state in the middle of the trap was observed. This experimental configuration is illustrated in Fig.~\ref{fig:snoke}. Another striking evidence of the relaxation mechanism modeled here is the very recent experimental result reported by Wertz {\em et al.}~\cite{wertz}. A detailed analysis of the latter experiments is under way.

In polariton condensates, a steady state can be reached when new particles are continuously injected by optical excitation. In order to avoid the pinning of the condensate phase by the excitation laser, the additional particles are created at a frequency different from the condensate one (nonresonant excitation). In the case of a far detuned laser, the excitation creates free electron hole pairs that relax to form an excitonic reservoir. The further relaxation of the excitons into the lower polariton branch then provides a gain for the condensate. A steady state is reached when gain and losses compensate each other. The stimulated part of the scattering into the lower polariton branch can be modeled by introducing a gain term in the Gross-Pitaevskii equation~\cite{nonresonant,keeling}. The full dynamics of the polariton gas including trapping, nonresonant pumping and relaxation is given by
\begin{multline}
i \frac{\partial}{\partial t} \psi=- \left[\frac{\hbar \nabla^2}{2m}+V_{\rm eff}-\frac{i}{2}[\gamma-R(n_R)]+g|\psi|^2 \right. \\
\left.+i\frac{\kappa \bar{n}}{2} 
\left(
\bar{\mu}- \frac{i \partial}{\partial t}
\right)
\right] \psi.
\label{eq:GPENR}
\end{multline}
Here $\gamma$ is the loss rate and the function $R(n_R)$ describes the gain of the lower polariton branch thanks to stimulated scattering from the nonresonantly excited excitons in the reservoir with density $n_R$. The exciton density can be described by the rate equation
\begin{equation}
\frac{d}{dt}n_R=-\gamma_R n_R-R(n_R) |\psi|^2+P,
\label{eq:dnr}
\end{equation}
where $\gamma_R$ is the reservoir relaxation rate and $P$ the nonresonant pumping rate.
The effective potential $V_{\rm eff}=V_{\rm trap}+V_{\rm exc}$ consists of the trapping potential $V_{trap}$ and the blue shift due to the high energy excitons, proportional to the pumping term $V_{\rm exc}=\mathcal G P$~\cite{shape}.

It is worth pointing out that Eq.~\eq{GPENR} actually describes two relaxation mechanisms: (i) from the excitonic reservoir into the lower polariton branch and (ii) from high to low energy polariton states. Only the second relaxation process is modeled with the new dissipation mechanism derived here [term on the second line of Eq.~\eq{GPENR}]. 

In the absence of the relaxation term, the model~\eq{GPENR},\eq{dnr} predicts a condensate at a single frequency [see Fig.~\ref{fig:snoke}(a)]. The frequency of the condensate coincides with the potential energy at the center of the pumping spot. Away from the pumping spot, the potential energy is converted into kinetic energy.

On the other hand, when the relaxation is included [Fig~\ref{fig:snoke}(b)], a condensate also appears at an energy close to the bottom of the trap. It is important to highlight that the steady state solution consists of two condensates at well defined frequencies. As our mean field model only describes the coherent part of the bosonic field, the condensate line width vanishes. 

Fig. 2 (b) is globally in agreement with the experimental observations in Ref.~\cite{snoke}. Differently from our simulations though, in experiments polariton luminescence was observed at all frequencies between the pump region and ground state~\cite{snoke}. Possibly microcavity disorder and spontaneous relaxation are responsible for these deviations.

Physically, it is likely that a large contribution to the relaxation within the lower polariton branch comes from the scattering of polaritons with reservoir excitons. Indeed, estimates based on the golden rule for exciton-phonon scattering~\cite{haug} yield a relaxation constant of the order of $\kappa\approx 10^{-4} \mume^2$. Our numerical simulations show relaxation toward the bottom of the lower polariton branch for $\kappa n\gtrsim 0.1$. If exciton-phonon scattering was the only relaxation mechanism, it would require a polariton density $n=10^{3} \mume^{-2}$, much larger than the experimental estimates~\cite{snoke}. 

Further applications of the relaxation in the description of polariton condensates include the relaxation of nonresonantly excited polariton condensates in periodic potentials~\cite{yamamoto,cerda}, in one dimensional polariton wires~\cite{wertz} and the interplay of phonon relaxation with parametric scattering~\cite{houdre-stevenson}. Our model may also be useful to describe the interaction of the thermal cloud with the condensed part of an equilibrium Bose gas at finite temperature~\cite{bose-finT}.

We are indebted to B. Pietka, I. Carusotto, K. Lagoudakis, T. Liew and F. Manni for stimulating discussions.

\end{document}